\documentclass[10pt,twocolumn,preprintnumbers,amsmath,amssymb,nofootinbib
,superscriptaddress]{revtex4-2}
\usepackage{hyperref}
\usepackage{graphicx,longtable,mathrsfs,color,array}

\usepackage[usenames,dvipsnames]{xcolor} % colour
\usepackage{amssymb,amsmath,mathtools,mathrsfs,slashed} % maths
\usepackage{epsfig,subfigure,placeins,float} % plots
\usepackage{booktabs,longtable,ctable,multirow} % tables
\usepackage{exscale,relsize} % scale 
\usepackage[normalem]{ulem} % editing
\usepackage{enumerate}
\usepackage{enumitem}
\usepackage{comment}
\usepackage{color}
\allowdisplaybreaks[1]

\newcommand{\be}{\begin{equation}}
\newcommand{\ee}{\end{equation}}
\newcommand{\bea}{\begin{eqnarray}}
\newcommand{\eea}{\end{eqnarray}}

\newcommand{\gr}{{\mbox{\tiny GR}}}
\newcommand{\rl}{{\mbox{\tiny R,L}}}
\newcommand{\pv}{{\mbox{\tiny PV}}}
\newcommand{\ppe}{{\mbox{\tiny ppE}}}
\newcommand{\cs}{{\mbox{\tiny CS}}}
\newcommand{\st}{{\mbox{\tiny ST}}}
\newcommand{\gf}{{\mbox{\tiny GF}}}

\begin{document}
\title{Parametrized Parity Violation in Gravitational Wave Propagation}
\author{Leah Jenks}
\email{ljenks@uchicago.edu}
\affiliation{Kavli Institute for Cosmological Physics,\\ University of Chicago,  Chicago, IL 60637, USA}
\author{Lyla Choi}
\email{lc3535@princeton.edu}
\affiliation{Department of Physics, Princeton University, Princeton, NJ 08544, USA}
\author{Macarena Lagos}
\email{m.lagos@columbia.edu}
\affiliation{Department of Physics and Astronomy, Columbia University, New York, NY 10027, USA}
\author{Nicol\'as Yunes}
\email{nyunes@illinois.edu}
\affiliation{Illinois Center for Advanced Studies of the Universe, Department of Physics, University of Illinois at Urbana-Champaign, Urbana, Illinois 61801, USA.}
%\author{...}
\date{Received \today; published -- 00, 0000}

\begin{abstract}
Gravitational parity violation arises in a variety of theories beyond general relativity. Gravitational waves in such theories have their propagation altered, leading to birefringence effects in both the amplitude and speed of the wave. In this work, we introduce a  generalized, theory-motivated parametrization scheme to study parity violation in gravitational wave propagation. This parametrization maps to parity-violating gravity theories in a straightforward way. We find that the amplitude and velocity birefringence effects scale with an effective distance measure that depends on how the dispersion relation is modified. Furthermore, we show that this generic parametrization can be mapped to the parametrized-post-Einsteinian (ppE) formalism with convenient applications to gravitational wave observations and model-agnostic tests of general relativity. We derive a mapping to the standard ppE waveform of the gravitational wave response function, and also find a ppE waveform mapping at the level of the polarization modes, $h_+$ and $h_\times$. Finally, we show how existing constraints in the literature translate to bounds on our new parity-violating parameters and discuss avenues for future analysis.

\end{abstract}

\date{\today}
\maketitle

%------------------------

%---------
%SECTION: INTRODUCTION
%---------
\section{Introduction}
\label{sec:Intro}
The observation of gravitational waves (GWs) by the LIGO-Virgo-Kagra collaboration (LVK)\cite{LIGOScientific:2018mvr,LIGOScientific:2020ibl,LIGOScientific:2021djp,LIGOScientific:2017zic} and the current era of GW astrophysics have provided a rich background on which to test gravity and study fundamental physics \cite{Yunes:2016jcc, Nair:2019iur}. Although the LVK observations have thus far not observed significant deviations from Einstein's theory of general relativity (GR) \cite{LIGOScientific:2018dkp, LIGOScientific:2019fpa, LIGOScientific:2020tif, LIGOScientific:2021sio}, there are myriad modified gravity theories that can lead to deviations from GR. Generally, these types of theories are motivated from a high-energy ultraviolet (UV) theory that can lead to small corrections to GR at low energies in an effective field theory (EFT) perspective (see e.g. \cite{Alexander:2021ssr}). 

Modified gravity theories can have a variety of effects on GWs, which can be characterized by modifications to the GW amplitude and/or its phase. These modifications can arise both in the GW generation and propagation, but in this work, we will focus solely on the latter. Real propagation effects in the phase arise from modified dispersion relations, characterized for example in \cite{Mirshekari:2011yq, Ezquiaga:2021ler,Ezquiaga:2022nak}, and have been constrained by LVK observations. Amplitude modifications typically arise from imaginary modifications to the dispersion relations, and they are more difficult to constrain observationally because of degeneracies with other GW parameters. 

While these effects are generic features of many modified gravity theories, in this study we will focus on parity-violating theories. Parity violation should here be understood as the lack of invariance of the action that characterizes a theory under a parity transformation, i.e.~under the inversion of the spatial triad in a properly adapted coordinate system. Parity-violating gravity theories modify the propagation of GWs in a way that specifically leads to amplitude and/or velocity birefringence. Birefringence denotes the phenomenon in which the right- and left-handed polarization modes evolve differently in their propagation. Specifically, amplitude (velocity) birefringence occurs when the imaginary (real) part of the phase evolves differently for the left- and right-polarization states. 

The most widely studied parity-violating theory is Chern-Simons (CS) gravity \cite{Lue:1998mq, jackiw, Alexander:2009tp}, which is characterized by a (four-dimensional) gravitational Chern-Simons term coupled to a dynamical field that is added to the Einstein-Hilbert action. The parity-violating effects of this theory arise when the field is a cosmological or background scalar (even under parity), because when this couples to the Pontryagin invariant (odd under parity) the scalar-gravitational Chern-Simons term in the action is parity violating. Chern-Simons gravity has been studied in a variety of contexts, including its effects on inflation in the early universe, on black hole spacetimes, and on the generation and propagation of GWs  (see e.g. \cite{Alexander:2007zg, Alexander:2007vt, Alexander:2008wi, Sopuerta:2009iy, cardosogualtieri, konnoBH, amarilla, chen, Garfinkle:2010zx, Molina:2010fb, Yagi:2013mbt, Loutrel:2018ydv, Delsate:2018ome, Wagle:2019mdq, Wagle:2021tam,Alexander:2007kv,Yunes:2010yf, Alexander:2017jmt, Yagi:2017zhb, Grumiller:2007rv,Shiromizu:2013pna, Yunes:2009hc, Yagi:2012ya,Maselli:2017kic,Silva:2020acr,Nakamura:2018yaw,Yagi:2013mbt,Nair:2019iur,Perkins:2021mhb,Alexander:2022avt, Loutrel:2022tbk}). A variety of extensions of Chern-Simons gravity also exist, including Palatini Chern-Simons \cite{Sulantay:2022sag}, torsional Chern-Simons \cite{Bombacigno:2022naf, Boudet:2022nub}, and Einstein-Axion-Chern-Simons \cite{Nojiri:2019nar}. A theory which includes both a Chern-Simons term and a Gauss-Bonnet term \cite{Kawai:2017kqt} has also been suggested. All of these theories induce similar parity-violating modifications to the propagation of GWs.  

Chern-Simons gravity is the most well-studied parity-violating modified gravity theory because it has concrete motivations from particle physics \cite{ALVAREZGAUME1984269} and string theory  \cite{Polchinski:1998rr, PhysRevLett.96.081301, Green:1987mn, Alexander:2004xd}, and is also the unique parity-violating metric theory that one can write down that is quadratic in the curvature and linear in the associated scalar field \cite{Alexander:2009tp}. However, if one relaxes the above assumptions, we can consider other parity-violating theories of interest. These include ghost-free scalar-tensor gravity \cite{Crisostomi:2017ugk, Nishizawa:2018srh, Zhao:2019xmm,  Qiao:2021fwi}, the Symmetric Teleparallel equivalent of GR \cite{Conroy:2019ibo}, and versions of Horava-Lifshitz gravity \cite{Horava:2009uw,Zhu:2013fja}, all of which  lead to parity violation. Parity-violating ghost-free scalar-tensor gravity \cite{Crisostomi:2017ugk} includes Chern-Simons gravity as a particular limit, but generalizes the theory to include additional parity-violating terms by including higher derivatives of the associated scalar field.  The parity-violating symmetric teleparallel equivalent of GR \cite{Conroy:2019ibo} is constructed in terms of a non-Riemannian formulation such that the Einstein-Hilbert action can be written in terms of the `non-metricity' tensor. As a non-metric theory, one is allowed to construct additional parity-violating extensions to GR than just the CS interaction. Lastly, Horava-Lifshitz gravity was first introduced as an extension of GR that is renormalizable and UV complete \cite{Horava:2009uw}. This theory breaks Lorentz invariance and can include higher-order parity-violating terms \cite{Zhu:2013fja}. Parity-violating Horava-Lifshitz extends beyond CS gravity by introducing higher dimensional operators that contain derivatives of the curvature. 

Significant amounts of work have gone into parametrizing deviations from GR in GWs, but little of this work has focused on parity-violating effects. A widely used formalism is the parametrized post-Einsteinian (ppE) framework, which treats corrections due to modifications of gravity in analogy to post-Newtonian (PN) corrections in GR \cite{PPE}. This parametrization scheme  can take into account both generation effects and propagation effects \cite{PhysRevD.57.2061, Mirshekari:2011yq, Chatziioannou:2012rf, Sampson:2013wia, cornishsampson} and has also been extended beyond simple binary geometries, see e.g., \cite{Huwyler:2014gaa,Huwyler:2011iq,Loutrel:2014vja, Mezzasoma:2022pjb, Loutrel:2022xok}. Separately from this, there has also been work studying waveform parametrizations of parity-violating gravity specifically \cite{Zhao:2019xmm, Zhao:2019szi, Qiao:2022mln}. These approaches are aimed at creating a generic GW template for the predictions of various parity-violating theories, but the non-GR corrections are constructed from complicated integrals that differ from theory to theory. Thus, the waveform cannot be studied generically and cannot be easily mapped to specific constraints on any particular theory. Furthermore, such non-GR corrections cannot be translated to a generic parametrization \textit{a la} ppE for the GW strain or in the $+/\times$ polarization basis. 

In addition to being a convenient theoretical method to characterize modified GW waveforms, the ppE formalism has also become an important tool to test GR with gravitational wave data. PpE corrections to the waveform amplitude or phase can enter at any post-Newtonian orders, and thus, it is straightforward to perform a full Bayesian parameter estimation study or Fisher forecasts. Previous work in the literature that analyzed gravitational wave data to search for parity violation used non-ppE methods for a limited number of waveform corrections and without a consistent parametrization. Amplitude birefringence was studied in \cite{Okounkova:2021xjv} and \cite{Ng:2023jjt} for one particular choice of a theory-independent parametrization using GW data from the GWTC-2 and GWTC-3 catalogs. Similarly, data analysis of velocity birefringence modifications with the GWTC-2 and GWTC-3 catalogs was performed in \cite{Wang:2020cub, Zhao:2022pun}, but for one particular parametrization of the parity-violating effects without a full ppE analysis. Thus, a full analysis of gravitational parity violation within the ppE framework has not yet been performed. 

Our aim in this work is to introduce a theory-motivated yet \textit{still} theory-agnostic framework that allows for straightforward theoretical studies of gravitational parity violation and applications to real world data analysis. We realize this aim by first introducing parametrized parity-violating modifications to the GW propagation equations. By working within the framework of an effective field theory, we present an explicit argument that shows that these modifications \textit{must be} the only parity-violating interactions one can construct. We show how the addition of these parity-violating corrections forces the right- and left-handed GW polarization modes to propagate differently, leading to amplitude and velocity birefringence. With the assumption that any parity-violating corrections are small deviations from GR and that any time evolution of parity-violating parameters is small compared to the expansion of the universe, we further derive explicit forms for the corrections to the amplitude and phase of the GW. Within the above-mentioned assumptions, these expressions for the amplitude and phase are exact, and we explicitly characterize the distance scaling of the parity-violating effects. 

We then show that our parametrized parity-violating waveforms map exactly to the ppE framework, and moreover, they both map exactly to known predictions of parity-violating theories studied in the literature. We calculate the modified waveform response function, and discuss degeneracies between the parity-violating modifications and waveform parameters. We then express the waveform modifications in terms of a ppE waveform, both at the level of the response function to the detector $\tilde{h}$ and the individual polarizations $h_+$ and $h_\times$, and show how the parity-violating ppE construction maps onto specific theories. The detailed mapping to ppE allows for straightforward data analysis, which we show explicitly by calculating how our parity-violating parameters are constrained by currently-existing analyses in the literature. Given that the existing analyses are limited in their scope, we discuss future directions to constrain parity-violation with future and existing GW data.

The structure of the paper is as follows. In Sec.~\ref{sec:genparam} we introduce our parametrization for parity-violating gravity theories and show how it leads to amplitude and velocity birefringence for a GW in the circular right- and left-handed (R/L) polarization basis. We also summarize how the parametrization maps onto specific theories. In Sec.~\ref{Sec:waveform} we show how the parametrization in the R/L polarization basis maps onto the waveform in the $+/\times$ basis, and we show that it can furthermore be mapped onto a ppE waveform. We then discuss existing constraints that can be applied to our new parity-violating parameters in Sec. \ref{Constraints}, and finally conclude with discussion and directions for future work in Sec.~\ref{Sec:discussion}. Throughout the remainder of this paper, we use the following conventions: we employ geometric units such that $G=1=c$ and assume a $(-, +, +, +)$ metric signature; Latin letters (a,b,...,h) range over all spacetime coordinates, (i,j,k) range over spatial indices, and square brackets denote anti-symmetrization over indices. 

\section{General parametrization for Parity-Violating GW Propagation}
\label{sec:genparam}

Our goal in this work is to find a theory-agnostic parametrization that can be used to study amplitude and velocity birefringence in GW propagation from binary black hole and neutron star events. To achieve this, we study relevant parity-violating extensions of GR and find the modifications to GW propagation in each theory. Then, we use these results to infer a generalized parametrization. We additionally show from a generic perspective why the parametrization must be the correct one. For clarity, we present the theory-agnostic result here, and enumerate details of specific theories in Appendix \ref{App:Theories}. 

We first introduce a general overview of parity-violating gravity in Sec.~\ref{PVGravity}. In Sec.~\ref{PPV}, we introduce a parity-violating parametrization of the GW propagation equations and show that it must be the correct and most general expression to characterize parity violation in GW propagation. In Sec.~\ref{GWProp}, we discuss how the GW propagation is modified and show how it leads to GW amplitude and velocity birefringence for parity-violating theories. Finally, we show the mapping to specific theories in Sec.~\ref{Sec:TheoryMap}. 

\subsection{Parity-Violating Gravity}
\label{PVGravity}

In general, one can describe the action of a parity-violating gravity theory as follows:
\be
S = S_\gr  + S_\pv ,
\ee 
where the Einstein-Hilbert action is
\be 
S_\gr  = \int d^4x \sqrt{-g} R,
\ee 
and $S_\pv $ is the beyond-GR parity-violating contribution to the action. We will not assume a specific form for $S_\pv $, but examples of explicit expressions for $S_\pv $ can be found in Appendix \ref{App:Theories} for the theories we have considered. In general, $S_\pv$ can be a function of the curvature and an auxiliary scalar field, $\varphi$. 

The addition of $S_\pv $ leads to a modification of the linearized field equations and thus to the GW dispersion relation. Let us start by considering a cosmological  Friedmann-Robertson-Walker (FRW) background, and linear perturbations around it. In general, a metric tensor could carry up to 6 different polarizations---two of each helicity 2, helicity 1 and helicity 0. However, due to the symmetries of the background, the linear perturbations of these different helicities decouple and their dynamics can be analyzed separately. In this paper, we focus on the dynamics of the helicity-2 polarizations. Additional polarizations present in the theories discussed here will be analyzed in the future.

Let us then proceed to consider the two helicity-2 polarizations of GWs and use a circular basis: right-handed and left-handed (R/L). On FRW, the linear perturbations of GWs in spatial Fourier space can be expressed as
\be 
h_\rl (\eta) = \mathcal{A}_\rl (\eta)e^{-i[\phi(\eta) - k_i x^i]},  \label{h_Fourier}
\ee 
where $\eta$ is conformal time, $d\eta = dt/a$ with $a$ the scale factor, $\phi(\eta)$ is the GW phase, $k_i$ is the comoving wavenumber vector, and $\mathcal{A}_\rl $ is the right and left-handed polarization amplitude, respectively. Notice that $h_\rl$ is complex in that it can be written in terms of an amplitude and a phase. Furthermore, we will see that $\phi(\eta)$ can have both real and imaginary contributions. The real part of $\phi$ will be the true phase of the GW, while the imaginary part will contribute to the overall amplitude. 

From here, the propagation equations of GWs that break parity symmetry but still preserve spatial rotation and translation invariance, and contain up to second-order time derivatives can generically be expressed as 
\begin{widetext}
\begin{align}\label{h_Eq}
h_\rl ^{''} &+ \left\{2\mathcal{H} +\lambda_\rl \sum_{n=1}^\infty k^n \left[\frac{\alpha_n(\eta)}{(\Lambda_\pv a)^n}\mathcal{H} + \frac{\beta_n(\eta)}{(\Lambda_\pv a)^{n-1}} \right]\right\}h'_\rl  \nonumber \\
&+ k^2 \left\{1 + \lambda_\rl \sum_{m=0}^\infty k^{m-1}\left[\frac{\gamma_m(\eta)}{(\Lambda_\pv a)^m}\mathcal{H} + \frac{\delta_m(\eta)}{(\Lambda_\pv a)^{m-1}}\right]\right\} h_\rl =0,
\end{align}
\end{widetext}

where $k=|\vec{k}|$ is the magnitude of the comoving wavenumber, primes denote derivatives with respect to conformal time, $\mathcal{H} := a'/a$ is the comoving Hubble parameter, and $\lambda_\rl  = \pm 1$. We take $n$ to be an odd, positive integer, and $m$ to be an even, non-negative integer, such that $m-1$ is odd. In this paper, we will consider parity-violating modifications within the context of a low-energy effective field theory, such that $\Lambda_\pv $ is the  cutoff scale of the theory. The dimensionless functions $\alpha, \beta, \gamma,$ and $\delta$ parametrize the magnitude of the parity violation. In the most general case, these functions depend on conformal time, as we have explicitly denoted above. This is due to the fact that parity-violating modified gravity theories generally propagate auxiliary fields, which themselves evolve with time, and thus, the coefficient in the effective-field-theory expansion can become complicated functions of these fields and their derivatives. For now, we will keep this time dependence general, but we will later make simplifying assumptions about the time evolution of these functions, such that they can be approximated by their values today. Note that within the EFT framework, we assume that all parity-violating modifications must be small, and we keep only the leading-order terms. As a consistency check, we see that, when $\alpha= \beta= \gamma=\delta=0$, we recover the propagation equation for GWs in GR. 

\subsection{Parametrized Parity-Violating Propagation Equations}
\label{PPV}
Let us pause momentarily to further motivate the form of Eq.~\eqref{h_Eq}. This expression was found by finding the parity-violating corrections to the GW amplitude and phase in multiple theories individually (see Appendix~\ref{App:Theories}), and then, inferring from the results a general form for the corrections that maps backwards to the individual form for the propagation equation. However, we can also see from a generic perspective that this form for the parity-violating modifications has to be correct by considering the following. 

As stated above, we know that the propagation equations can be up to second order in time derivatives, which allows for three possible modifications, to the dispersion relation: one each on the functions that multiply $h_\rl ^{''}, h_\rl ',$ and $h_\rl $:
\begin{align}
    A \, h_\rl ^{''} + B \, h_\rl ' + C \, h_\rl  = 0\,, \label{eq:genmod}
\end{align}
where $A, B$ and $C$ are functions of time (through the scale factor and its derivatives) and of $k$, and may be different for right- or left-handed polarizations. Since the left-hand side of this equation equals zero, we can divide by A and rewrite the equation in a simpler form:
\begin{align}
    h_\rl ^{''} + \bar{B} \, h_\rl ' + \bar{C} \, h_\rl  = 0\,,\label{eq:genmod2}
\end{align}
where $\bar{B}$ and $\bar{C}$ are  new functions of, again, time and $k$, and possibly different for  right- and left-handed polarizations. 

Let us now determine the functional dependence of these functional prefactors by considering the properties of propagating GWs and using dimensional analysis. The only physical quantities that are able to influence GW propagation in parity-violating theories are the frequency of the wave, the distance that the wave has propagated over, and the magnitude and time evolution of the coupling parameter(s) that encode parity violation. The distance of propagation can be replaced with the Hubble constant today times some function of redshift (or analogously by the Hubble parameter). The coupling parameters can be product-factorized into a dimensionless coupling constant(s) and a cut-off scale $\Lambda_{\pv}$ to a given power. Moreover, by dimensional analysis, $\bar{B}$ and $\bar{C}$ have to have units of inverse conformal time to the first and second power respectively. The only quantities that have units of inverse conformal time are precisely the three mentioned above: $k$, ${\cal{H}}$ and $\Lambda_\pv a$. Notice also that because the coupling parameters are allowed to be time-varying, one can construct an arbitrary function of them and of their derivatives that also has the correct units of inverse conformal time. We will call this function $f(\varphi')$, where we use $\varphi$ to make contact with the fact that additional degrees of freedom in most parity-violating modified gravity theories are scalar or pseudo-scalar fields, such as in Chern-Simons theory. 

Using all of this, we can infer that 
\begin{align}
    \bar{B} &= k \, \bar{B}_k + {\cal{H}} \, \bar{B}_{\cal{H}} + \Lambda_{\pv} a \, \bar{B}_{\Lambda} \, + f(\varphi') \, 
    \bar{B}_\varphi \,,\label{eq:BarBC}
    \\
    \bar{C} &= k^2 \bar{C}_{kk} + \mathcal{H}^2\bar{C}_{\mathcal{H} \mathcal{H}} + (\Lambda_\pv a)^2 \bar{C}_{\Lambda \Lambda} \nonumber \\
    &+ k\mathcal{H} \bar{C}_{k\mathcal{H}} + k\Lambda_\pv a \bar{C}_{k\Lambda} + \mathcal{H}\Lambda_\pv a \bar{C}_{\mathcal{H}\Lambda}\nonumber \\
    &+ kf(\varphi')\bar{C}_{k\varphi} + f(\varphi')\Lambda a \bar{C}_{\varphi \Lambda} + f(\varphi')\mathcal{H}\bar{C}_{\mathcal{H}\varphi}.\label{eq:BarC}
\end{align}
where $\bar{B}_i$ and $\bar{C}_{ij}$ with $i,j \in (k, {\cal{H}}, \Lambda_\pv a, \phi)$ are now dimensionless functions. These functions must depend on dimensionless combinations of $k$, ${\cal{H}}$, $\Lambda_{\pv} a$ and $f(\varphi')$, and thus, 
\begin{align}
 \bar{B}_i &= \bar{B}_i\left(\frac{k}{\Lambda_{\pv} a},\frac{\mathcal{H}}{\Lambda_{\pv} a}, \frac{f(\varphi')}{\Lambda_{\pv}a} \right)\,,\label{Bi}\\
   \bar{C}_{ij} &= \bar{C}_{ij}\left(\frac{k}{\Lambda_{\pv} a},\frac{\mathcal{H}}{\Lambda_{\pv} a}, \frac{f(\varphi')}{\Lambda_{\pv}a}\right).\label{Cij}
\end{align}
In principle, further dimensionless combinations could appear in Eqs.~\eqref{Bi} and \eqref{Cij}, which contain $k, \mathcal{H}$ and $f(\varphi')$ in the denominator, for example $\frac{\mathcal{H}}{k}$ or $\frac{k}{f(\varphi')}$. However, the functions $\bar{B}_{i}$ and $\bar{C}_{ij}$ must be finite as $\mathcal{H}\rightarrow 0$, $k\rightarrow 0$, and $f(\varphi') \rightarrow 0$ in order to ensure a continuous parametrization connected to flat space, low energy, and GR, respectively. Thus, these terms must not be present, and we are left only with the dimensionless quantities appearing in Eqs.~\eqref{Bi} and \eqref{Cij}.

With this in hand, we can express the dimensionless functions as 
\begin{align}
    \bar{B}_i  &= \sum_{n=0,m=0}^\infty b_{nm}^{(i)} (\eta)\left(\frac{k}{\Lambda_{\pv} a}\right)^n \left(\frac{{\cal{H}}}{\Lambda_{\pv} a}\right)^m\,,\label{eq:Bi}\\
     \bar{C}_{ij}  &= \sum_{n=0,m=0}^\infty c_{nm}^{(ij)}(\eta) \left(\frac{k}{\Lambda_{\pv} a}\right)^n \left(\frac{{\cal{H}}}{\Lambda_{\pv} a}\right)^m\,,\label{eq:Ci}
\end{align}
where $b_{nm}^{(i)}$ and $c_{nm}^{(ij)}$ are coefficients that are allowed to vary in time. The sum on $m$ cannot start at a negative number because, as we mentioned above, these functions must be regular in the flat spacetime limit. The sum on $n$ must also start at positive values (or zero) because the deviations from GR must decay as the cut-off is increased. 

Let us now  simplify several of the terms in Eqs.\ (\ref{eq:BarBC})-(\ref{eq:BarC}). Notice that with the expansion in Eqs.~\eqref{eq:Bi} and ~\eqref{eq:Ci}, several terms are degenerate with each other and can be combined. For example, all terms in the $\bar{B}_k$ sums in Eq.~\eqref{eq:Bi}can be represented with different $m$ and $n$ choices in the $\bar{B}_\Lambda$ and $\bar{B}_\mathcal{H}$ sums. For example, if we consider the $n=1$ and $m=0$ term in the $\bar{B}_k$ sum, we obtain $k\bar{B}_k = k^2/(\Lambda_\pv a)$; taking $n=2$ and $m=0$ in the $\bar{B}_\Lambda$ sum gives $\Lambda_\pv B_\Lambda = k^2/(\Lambda_\pv a)$. Similarly, $k\bar{B}_k(n=0,m=1) = \mathcal{H}\bar{B}_\mathcal{H}(n=1, m=0)$. Notice also that the time evolution induced by the $f(\varphi')\bar{B}_{\varphi}$ term can be absorbed into $b_{nm}^\Lambda(\eta)$ and $b_{nm}^\mathcal{H}(\eta)$. Thus, all contributions can be represented with the second and third terms in Eq.~\eqref{eq:BarBC}. We can perform a similar analysis for the $\bar{C}_{ij}$ terms. The three terms proportional to $\bar{C}_{k\mathcal{H}}, \bar{C}_{k\Lambda}$ and $\bar{C}_{\mathcal{H}\Lambda}$ can all be represented by the $\bar{C}_{kk}$ terms with particular choices of $n$ and $m$. Moreover, the $\bar{C}_{\mathcal{H} \mathcal{H}}$ term can be neglected because it is $\mathcal{O}(\mathcal{H})^2$, and again the time evolution from the $f(\varphi')$ terms can be absorbed into $c_{nm}^{(ij)}(\eta)$. Thus, we are left with only the $\bar{C}_{\Lambda \Lambda}$ and $\bar{C}_{kk}$ terms.

We have so far considered generic deformations of the GW propagation equation, but we would like to focus on modifications that specifically violate parity. The equations are said to preserve parity symmetry when by making a reflection of the spatial coordinates $\vec{x}\rightarrow -\vec{x}$ the physical solution does not change. More concretely, this means that both right- and left-handed polarizations have the same physical behavior, and hence when we make the exchange of polarizations $h_{L}\leftrightarrow h_{R}$, the equations stay the same because $\bar{B}$ and $\bar{C}$ are the same for right- and left-handed polarizations  in Eq.\ (\ref{eq:genmod2}).
On the contrary, if we have parity breaking equations, then $\bar{B}_{L}\not= \bar{B}_R$ and $\bar{C}_{L}\not= \bar{C}_R$. Nevertheless, the right- and left-handed coefficients are expected to be related to each other since typical theories studied in the literature come from an action principle, which is a scalar parity-preserving quantity. As such, it can only lead to a set of equations that satisfy the relations  $\bar{B}_{L,R}(\vec{k})=\bar{B}_\rl (-\vec{k})$ and $\bar{C}_{L,R}(\vec{k})=\bar{C}_\rl (-\vec{k})$. From here, we can then expect the only difference between right and left coefficients in parity-breaking theories to be a change in sign.
This implies specifically that the dimensionless functions $\bar{B}_i \leftrightarrow -\bar{B}_i$ and $\bar{C}_{ij} \leftrightarrow -\bar{C}_{ij}$ when $h_R \leftrightarrow h_L$. Let us then introduce the quantity $\lambda_\rl  = \pm 1$ for right and left, respectively. GW modifications that are parity-violating should be linear in $\lambda_\rl $. We can incorporate this effect explicitly by expanding the coefficients $b^{(i)}_{nm}$, assuming that any parity-violating effects are small perturbations from GR: 
\begin{align} 
b^{(i)}_{n,m} &= \bar{b}^{(i)}_{nm} + (\lambda_\rl )^n(\lambda_\rl )^m\delta b^{(i)}_{nm},\\
c^{(ij)}_{nm} &= \bar{c}^{(ij)}_{nm} + (\lambda_\rl )^n(\lambda_\rl )^m\delta c^{(ij)}_{nm} \,,
\end{align} 
where $\bar{b}_{nm}^{(i)}$ and $\bar{c}_{nm}^{(ij)}$ are the background GR quantities and the $\delta b_{nm}^{(i)}$ and $\delta c_{nm}^{(ij)}$ coefficients characterize the parity-violating corrections. 

We can now implement the above in the functional prefactors. Expanding out the relevant terms in Eq.~\eqref{eq:BarBC}, we have 
\begin{align}
\bar{B} &= \sum_{n=0,m=0}^\infty \Bigg[\mathcal{H}\left(\bar{b}^{\mathcal{H}}_{nm} + \lambda_\rl ^{n+m}\delta b^{\mathcal{H}}_{nm}\right)\left(\frac{k}{\Lambda_{\pv} a}\right)^n \left(\frac{{\cal{H}}}{\Lambda_{\pv} a}\right)^m, \nonumber \\
 &+ \Lambda_\pv a \left(\bar{b}^\Lambda_{nm} + \lambda_\rl ^{n+m}\delta b^\Lambda_{nm}\right)\left(\frac{k}{\Lambda_{\pv} a}\right)^n \left(\frac{{\cal{H}}}{\Lambda_{\pv} a}\right)^m\Bigg].
\end{align}
First, let us delineate which terms arise from GR in the usual expression for GW propagation and which ones arise from parity-violating EFT considerations. From the GR expression, we must have that $\bar{b}^\mathcal{H}_{00}= 2$, $\bar{b}_{nm} =0$ for all $(n,m) \neq (0,0)$, and $\bar{b}^\Lambda_{nm}=0$ for all $n$ and $m$. Now consider the perturbative contributions. We assume that $k \gg \mathcal{H}$, e.g., that GW wavelengths are short compared to the Hubble scale, and thus will keep terms that are at most linear in $\mathcal{H}$, but that may contain higher powers of $k$. To obtain terms that are up to linear order in $\mathcal{H}$, we must have $\delta b^\mathcal{H}_{nm} =0$ for $m\neq0$, so we are left with $\delta b^{\mathcal{H}}_{n0}$ nonvanishing. Terms with $m\neq 0$ are degenerate with terms arising from the $\mathcal{H}\bar{B}_\mathcal{H}$ expansion, and thus, we are left with $\delta b^\Lambda_{10} \neq 0$. We also see that for $n$ even, $(\lambda_\rl )^n = +1$ and the modifications will be parity preserving. Thus, we require that $n$ be an odd integer. 

Similarly, for $\bar{C}$, we have:
\begin{align}
    \bar{C} &=\sum_{n=0,m=0}^\infty \Bigg[ k^2 \left(\bar{c}^k_{nm} + \lambda_\rl ^{n+m}\delta c_{nm}^k\right)\left(\frac{k}{\Lambda_{\pv} a}\right)^n \left(\frac{{\cal{H}}}{\Lambda_{\pv} a}\right)^m,\nonumber \\
    &+ (\Lambda_\pv a)^2 \left(\bar{c}^\Lambda_{nm} + \lambda_\rl ^{n+m}\delta c_{nm}^\Lambda\right)\left(\frac{k}{\Lambda_{\pv} a}\right)^n \left(\frac{{\cal{H}}}{\Lambda_{\pv} a}\right)^m\Bigg].
\end{align}
From GR, we have that $\bar{c}_{00}^k = 1$ with $\bar{c}_{nm}^k =0$ for all $(n,m) \neq (0,0)$. We can see that $\delta c^\Lambda_{00},\delta c^\Lambda_{10},$ and $\delta c^\Lambda_{01}$ all must vanish as they increase as $\Lambda_\pv$ increases. To keep the expression linear in $\mathcal{H}$, $\delta c^\Lambda_{nm}$ with $m > 1$ can be neglected, so we are left with $\delta c^\Lambda_{n1}$ nonvanishing for $n > 1$. Finally, consider $\delta c^k_{nm}$. We find that $\delta c_{n0}^k$ is nonvanishing, while the $\delta c_{01}^k$ contribution is degenerate with the $\delta c_{31}^\Lambda$ contribution and again $\delta c_{nm}^k$ with $m > 0$ can be neglected at this level. 
 
With this analysis, we are left with 
\begin{align}
    \bar{B} &= 2\mathcal{H} \nonumber \\
    &+ \lambda_\rl \sum_{n=1}^\infty \Bigg[\delta b_{n0}^\mathcal{H}\mathcal{H}\left(\frac{k}{\Lambda_{\pv} a}\right)^n + \delta b^\Lambda_{n0}\Lambda_\pv a \left(\frac{k}{\Lambda_{\pv} a}\right)^n  \Bigg], \\
    \bar{C} &=  k^2  \nonumber \\
    &+\lambda_\rl \sum_{n=1}^\infty\Bigg[\mathcal{H}\Lambda_\pv a \delta c^\Lambda_{n1} \left(\frac{k}{\Lambda_{\pv} a}\right)^n + \delta c_{n0}^k k^2 \left(\frac{k}{\Lambda_{\pv} a}\right)^n\Bigg],
\end{align}
where $n$ is odd. We see that $\bar{B}$ and $\bar{C}$ give us expressions for the parity-violating modifications exactly as we have in Eq.~\eqref{h_Eq}, identifying $\alpha_n(\eta) = \delta b^\mathcal{H}_{n0}$, $\beta_n(\eta) = \delta b^\lambda_{n0}$, $\gamma_m(\eta) = \delta c^\Lambda_{nm}$, and $\delta_m(\eta) = \delta c_{n0}^k$. 

Let us summarize the assumptions we have made to obtain the above expression in the following list:
\begin{enumerate}
    \item Any deviation from GR is small, and the modifications can be reliably treated in an EFT framework.
    \item The beyond-GR modifications are specifically parity violating. Other non-parity-violating terms could appear with even powers of $\lambda_\rl  k$, but as these types of modifications have been more extensively studied in the literature, we focus solely on the parity-violating contributions. 
    \item We expect the modifications to be polynomial in $k$, assuming both locality and small deviations from GR. Generically, one could expect rational polynomials in $k$, because $A, B,$ and $C$ in Eq.~\eqref{eq:genmod} could all be polynomials. However, in assuming small deviations from GR, we recover a simple polynomial form for $\bar{B}$ and $\bar{C}$. 
    \item We assume that $k \gg \mathcal{H}$, and that the GW wavelengths are short compared to the expansion of the universe.
\end{enumerate} 
Thus, with the above discussion, we see that Eq.~\eqref{h_Eq} describes the most generic parity-violating corrections to GW propagation allowable, within the above listed assumptions.

\subsection{Modified GW Propagation}
\label{GWProp}
Replacing Eq.\ (\ref{h_Fourier}) into Eq.\ (\ref{h_Eq}) we can rewrite the GW equation in terms of the phase only, and can write an effective modified dispersion relation as 
\begin{align}
    \phi^{''} &+ \phi'^2 + i\phi'\left\{2\mathcal{H} + \lambda_\rl k^n \left[\frac{\alpha_n}{(\Lambda_\pv a)^n}\mathcal{H} + \frac{\beta_n}{(\Lambda_\pv a)^{n-1}} \right]\right\} \nonumber \\
    &- k^2 \left\{1 + \lambda_\rl  k^{m-1}\left[\frac{\gamma_m(\eta)}{(\Lambda_\pv a)^m}\mathcal{H} + \frac{\delta_m(\eta)}{(\Lambda_\pv a)^{m-1}}\right]\right\} =0,\label{phi_PV_eqn}
\end{align}
where we have kept the sums over $n$ (odd) and $m$ (even) implicit, and we have assumed that the GW amplitude varies on much longer timescales than the phase. Now, assuming that the parity-violating terms are small perturbations from GR, we can linearize the equations of motion by taking $\phi = \bar{\phi} + \delta\phi$, where the background solution $\bar{\phi}$ has the usual GR form, $\bar{\phi}' = \pm k - i\mathcal{H}$. Note that the $i\mathcal{H}$ contribution to $\bar{\phi}$ simply induces a decay in the GW amplitude with the expansion of the universe. This is accounted for in the standard form for the GW waveform signal, which includes a decay factor of $1/D_L$, where $D_L$ is the luminosity distance. In addition, $\delta\phi$ has both real and imaginary parts associated to velocity birefringence and amplitude birefringence contributions, respectively. We denote this as  
\be 
\delta\phi = -i\lambda_\rl \delta\phi_A + \lambda_\rl \delta\phi_V,
\ee 
where $A,V$ denote the amplitude and velocity contributions, respectively. We replace $\phi = \bar{\phi} + \delta\phi$ into Eq.\ (\ref{phi_PV_eqn}) and perform a series expansion assuming that $\delta\phi \ll \bar{\phi}$, ${\phi}^{''} \ll (\phi')^2$, and $\delta\phi^{''} \ll\bar{\phi}\delta\phi'$ \cite{Yunes:2010yf}. We then find for $\delta\phi'_{A,V}$:
\begin{align}
\delta\phi_A' &=  \frac{k^n}{2}\left[\frac{\alpha_n}{(\Lambda_\pv a)^n}\mathcal{H} + \frac{\beta_n}{(\Lambda_\pv a)^{n-1}} \right],\\
\delta\phi_V' &= \frac{k^m}{2}\left[\frac{\gamma_m}{(\Lambda_\pv a)^m}\mathcal{H} + \frac{\delta_m}{(\Lambda_\pv a)^{m-1}}\right].
\end{align} 

To simplify these expressions further, we will make the assumption that the time-dependent parity-violating parameters $\alpha, \beta, \gamma$, and $\delta$ vary slowly with respect to the expansion of the universe, and can thus be well approximated by their current values via a Taylor expansion, e.g., $\alpha_n \approx \alpha_{n_0}$. The comoving wavenumber $k$ is a constant in conformal time and can also be taken out of the integral. Additionally, we will employ the transformation $dt = -{dz}/[{H(z)(1+z)}]$. Then, using all of the above, we can rewrite the expressions for the phase and amplitude shifts as 
\begin{align}
    \delta\phi_A &=  \frac{k^n}{2} \left[\frac{\alpha_{n_0}}{\Lambda_\pv ^n}\int \frac{dz}{(1+z)^{1-n} } +\frac{\beta_{n_0}}{\Lambda_\pv ^{n-1}} \int\frac{dz}{H(1+z)^{1-n}} \right],\\
    \delta\phi_V &= \frac{k^m}{2}\left[\frac{\gamma_{m_0}}{\Lambda_\pv ^m}\int \frac{dz}{(1+z)^{1-m} } +\frac{\delta_{m_0}}{\Lambda_\pv ^{m-1}} \int\frac{dz}{H(1+z)^{1-m}} \right].
\end{align}

We can now define an effective distance, $D_\alpha$ as in \cite{Mirshekari:2011yq, Ezquiaga:2022nak},
\be 
D_\alpha = (1 +z)^{1-\alpha}\int \frac{(1 + z)^{\alpha-2}}{H(z)}dz,
\ee 
and will also in analogy define an effective redshift parameter, $z_\alpha$, such that 
\be 
z_\alpha = (1+z)^{-\alpha}\int \frac{dz}{(1 + z)^{1 - \alpha}}.
\ee 
Notice that $D_1 = D_T$, where $D_T$ is the look-back distance, and $D_2 = (1+z)^{-1}D_C = D_A$, where $D_C$ and $D_A$ are the comoving and angular-diameter distances, respectively. From the effective redshift parameter, we obtain that $z_0 = \ln(1+z)$ and $z_1 = (1+z)^{-1}z$. 

With these definitions, we can summarize the parametrized parity-violating deviations from GR as follows:
\begin{align}
    \delta\phi_A &= \frac{\left[k(1+z)\right]^n}{2}\left(\frac{\alpha_{n_0}}{\Lambda_\pv ^n}z_n + \frac{\beta_{n_0}}{\Lambda_\pv ^{n-1}} D_{n+1}\right), \label{Eq:phiA}\\
\delta\phi_V &=\frac{\left[k(1+z)\right]^m}{2}\left( \frac{\gamma_{m_0}}{\Lambda_\pv ^m}z_m + \frac{\delta_{m_0}}{\Lambda_\pv ^{m-1}}D_{m+1}\right),
\label{Eq:phiV}
\end{align}
such that the right and left-handed polarization modes are modified in the following way
\begin{align}
h_\rl  &= \bar{h}_\rl \exp\left\{\mp\frac{\left[k(1+z)\right]^n}{2}\left(\frac{\alpha_{n_0}}{\Lambda_\pv ^n}z_n + \frac{\beta_{n_0}}{\Lambda_\pv ^{n-1}} D_{n+1}\right)\right\}\nonumber \\
&\times \exp\left\{\pm i\frac{\left[k(1+z)\right]^m}{2}\left( \frac{\gamma_{m_0}}{\Lambda_\pv ^m}z_m + \frac{\delta_{m_0}}{\Lambda_\pv ^{m-1}}D_{m+1}\right)\right\},
\label{hRLmod}
\end{align}
where $\bar{h}_\rl $ is the usual GR expression for the right and left-handed modes. 

From the parametrized dispersion relation, one can also immediately see how the GW velocity is modified in parity-violating theories. The group and phase velocities of a GW are given by $v_g = d\omega/dk$ and $v_p = \omega/k$, respectively, where $\omega$ for the right and left-handed polarizations satisfies the following dispersion relation: 
\be 
\omega^2_\rl  =k^2 \left\{1 + \lambda_\rl  k^{m-1}\left[\frac{\gamma_m(\eta)}{(\Lambda_\pv a)^m}\mathcal{H} + \frac{\delta_m(\eta)}{(\Lambda_\pv a)^{m-1}}\right]\right\} .
\ee 
Again, assuming that all parity-violating parameters are small deviations from GR, we find for the modified group and phase velocities, respectively:
\begin{align}
v_g^\rl  &= 1 + \frac{\lambda_\rl }{2}mk^{m-1} \left[\frac{\gamma_{m}}{(a\Lambda_\pv )^m} \mathcal{H} + \frac{\delta_{m}}{(a\Lambda_\pv )^{m-1}}\right], \label{Eq:vg} \\ 
v_p^\rl  &= 1 + \frac{\lambda_\rl }{2}k^{m-1} \left[\frac{\gamma_{m}}{(a\Lambda_\pv )^m} \mathcal{H} + \frac{\delta_{m}}{(a\Lambda_\pv )^{m-1}}\right]. \label{Eq:vp}
\end{align}
Clearly then, the right-handed and left-handed waves may have different propagation speeds. Phenomenologically, this means that, for long propagation times, the two polarizations may \textit{decohere} in time, and lead to two separate ``echoes'' of the emitted GW, one purely right-handed and another purely left-handed. This parity-violating decoherence has been discussed in \cite{Ezquiaga:2021ler}. For short propagation times, the two polarizations propagate coherently in time. In that case, modifications to the GW speed have been tightly constrained by the coincident GW and gamma ray burst event, GW170817. We will discuss applications of this constraint to these parametrized velocities in Sec.~\ref{Constraints}.   

\subsection{Mapping to Parity-Violating Theories}
\label{Sec:TheoryMap}
Now that we have established the generic parametrization above, we can map the parity-violating coefficients to various theories. This mapping is shown below in Table \ref{tab:TheoryMap}. The parameters listed in the table correspond to specific parameters defined in each theory. We show the mapping to Chern-Simons gravity \cite{Alexander:2007kv, Yunes:2010yf, Alexander:2017jmt} and its extensions Palatini Chern-Simons \cite{Sulantay:2022sag} and generalized Chern-Simons \cite{Nojiri:2019nar}. We also consider parity-violating ghost-free scalar-tensor gravity \cite{Crisostomi:2017ugk, Nishizawa:2018srh, Zhao:2019xmm,  Qiao:2021fwi}, both second and fourth order constructions of the parity-violating extensions to the Symmetric Teleparallel equivalent of GR \cite{Conroy:2019ibo}, and parity-violating Horava-Lifshitz gravity \cite{Zhu:2013fja}. 
Details of each theory, including the propagation equations and definitions of the particular parameters, can be found in Appendix \ref{App:Theories}. 

\begin{widetext}
\begin{center}
\begin{table}[h]
    \centering
    \begin{tabular}{|c|c|c|c|c|c|} \hline 
        Theory &$\alpha_1$ & $\beta_1$ & $\gamma_0$ & $\delta_2$ & $\delta_4$  \\ \hline
        \hline
        (Palatini \cite{Sulantay:2022sag}) Chern-Simons \cite{Alexander:2007kv, Yunes:2010yf, Alexander:2017jmt} &$\frac{4\tilde{\alpha}^\cs\dot{\vartheta}}{\kappa} $ & 0 & 0 & 0  & 0\\
        Generalized Chern-Simons\cite{Nojiri:2019nar}  & $24 \tilde{U}_0$ & 0 & 0 & 0 & 0 \\ 
        \hline
        Ghost-Free Scalar Tensor \cite{Nishizawa:2018srh} &$ \alpha^\gf $
 &$-\dot{\tilde{\alpha}}^\gf $
 & 0 & $(\alpha^\gf - \beta^\gf) $  & 0 \\
 \hline
        Symmetric Teleparallel I \cite{Conroy:2019ibo} &  0 &  0 & $ -4\alpha^\st$ & 0 & 0 \\ 
        \hline
        Symmetric Teleparallel II \cite{Conroy:2019ibo} & $(3\beta_2 - 3\beta_3) $ & $\dot{\tilde{\beta}}_2 $ & 0 & $(\beta_1 - \beta_3)$ & 0\\
        \hline
        Horava-Lifshitz \cite{Zhu:2013fja} & 0 & 0 &0 & $ -\alpha_1 $ & $\alpha_2$    
    \\ \hline
    \end{tabular}
    \caption{Dimensionless GW birefringence parametrizations for various modified gravity theories. Definitions of the specific parameters can be found in Appendix~\ref{App:Theories}. Notice that Palatini Chern-Simons, Chern-Simons and Generalized Chern-Simons make the same predictions at leading order beyond GR. }
    \label{tab:TheoryMap}
\end{table}
\end{center}
\end{widetext}

From Table \ref{tab:TheoryMap} we can see that when considering parity-violating theories beyond GR in the literature, the only nonvanishing coefficients within our assumptions are $\alpha_1, \beta_1, \gamma_0, \delta_2,$ and $\delta_4$. In this case, the modifications to the phase can be determined explicitly to give
\begin{align}
    \delta\phi_A &= \frac{k(1+z)}{2}\left(\frac{\alpha_{1_0}}{\Lambda_\pv }z + \beta_{1_0}  D_{2}\right),\label{Amp_Bire_Linear}\\
\delta\phi_V &=\frac{1}{2}\Bigg\{\gamma_{0_0}\ln(1+z) + \left[k(1+z)\right]^2\frac{\delta_{2_0}}{\Lambda_\pv }D_{3}\nonumber \\&+ \left[k(1+z)\right]^4\frac{\delta_{4_0}}{\Lambda_\pv ^{3}}D_{5}\Bigg\}.
\end{align}
In principle, one could construct other theories that lead to contributions from additional $n$ terms, but for now we focus on the nonzero coefficients discussed above. 

In extensions of CS gravity, there are additional terms that arise beyond those presented above, which can give rise to GW velocity modifications and amplitude modifications \cite{Sulantay:2022sag,Nojiri:2019nar}. However, these velocity modifications come with higher powers of $\mathcal{H}$ or of the parity-violating  parameter, and thus, are subdominant to the leading terms considered here. There is additionally a formulation of Chern-Simons gravity that includes torsion \cite{Bombacigno:2022naf}, where GWs have also been shown to exhibit birefringent behavior \cite{Boudet:2022nub}. This study, however, was carried out in the adiabatic limit and on a de Sitter background. Thus, we do not consider torsional CS in our theory map, as we would like to consider the most general GW propagation on the FRW spacetime. 
Lastly, there are additional terms in the fourth-order symmetric teleparallel equivalent of GR and a higher-order construction of ghost-free scalar-tensor gravity, discussed in \cite{Zhao:2019xmm}, but these also do not lead to additional corrections at the level of our analysis for the reasons discussed above. 

\section{Parity-Violating Waveform Modifications}
\label{Sec:waveform}
We have so far considered parity-violating GW modifications in the circularly polarized right- and left-handed basis. However, to connect with GW observations, in this section we discuss how parity-violating modifications in the circular basis impact GWs in the linear polarization ($+/\times$) basis, commonly used in the literature, and how such modifications enter the waveform. More specifically, in this section we show how the modified $h_\rl $ in Eq.~\eqref{hRLmod} translates to a modified $h_{+, \times}$ and a modified response function, $h$. We then show how both of these modifications can be directly mapped to the ppE formalism \cite{PPE}. 

\subsection{Linearly Polarized Gravitational Waves}
The linear $+/\times$ polarization modes are defined in terms of the circular right- and left-handed polarization modes as 
\begin{align}
h_{+} &= \frac{h_R + h_L}{\sqrt{2}},\\
h_\times &= i \frac{h_R - h_L}{\sqrt{2}}.
\end{align}

Again, recalling that $\delta\phi_A$ and $\delta\phi_V$ are small deviations from GR, we can expand the exponential dependence of the parity-violating waveform modifications in Eq.~\eqref{hRLmod}. With this assumption we find that the polarization modes in the + and $\times$ basis are 
\begin{align}
h_+ &= \bar{h}_+ - i\delta\phi_A \bar{h}_\times + \delta\phi_V \bar{h}_\times, \label{hp}\\
h_\times &= \bar{h}_\times + i\delta\phi_A \bar{h}_+ - \delta\phi_V \bar{h}_+ \label{hx},
\end{align}
where $\bar{h}_{+,\times}$ are linearly polarized GWs in GR.
Notice that these modifications have been analyzed in spatial Fourier space in the previous sections. However, GWs are typically analyzed either in temporal Fourier space (i.e.~ in the ``frequency domain'') or in the time domain, since they are detected at a fixed location but at different times.  

The equations presented in the previous sections allow us to propagate the GW signal from the source to the detector in $k$ space. The observed signal in $k$-space can then be translated to that  in $f$ space by using, once more, that the local deviations from GR are small, and, thus, we have that $\omega\approx ck$ at the detector (although the accumulated deviations over the entire propagation time may not be negligible). We can thus simply replace  $k\rightarrow  2\pi f$ with $f$ being the detected GW frequency and $h_{+,\times}(k,\eta)\rightarrow \tilde{h}_{+,\times}(f,\eta)$ in the solutions previously calculated. Notice that $f$ is the detected and redshifted frequency, which is related to the emitted source frequency $f_s$ by $f_s=(1+z)f$. Therefore, the GR waveform in Eqs.\ (\ref{hp})-(\ref{hx}), denoted by $\tilde{\bar{h}}_{+,\times}$ in frequency space, is also a function of the redshifted quantities, such as $f$ and the redshifted chirp mass for compact binary coalescence.

We then calculate the measured GW response function to a detector, $\tilde{h}$ in the frequency domain in terms of the $\tilde{h}_{+,\times}$ modes as
\be 
\tilde{h} = F_+\tilde{h}_+ + F_\times \tilde{h}_\times,\label{strain}
\ee 
where $F_{+,\times}$ are the detector beam pattern functions, which depend on the GW source sky location and polarization angle. For a LIGO/Virgo type of L-shaped interferometer, they can be explicitly written as \cite{Sathyaprakash:2009xs}
\begin{align}
        F_+ &= \frac{1}{2}[1+\cos^2(\theta)]\cos(2\phi)\cos(2\psi)\nonumber\\
        &-\cos(\theta)\sin(2\phi)\sin(2\psi), \label{Fp} \\
    F_\times&= \frac{1}{2}[1+\cos^2(\theta)]\cos(2\phi)\sin(2\psi)\nonumber\\
    &+\cos(\theta)\sin(2\phi)\cos(2\psi),
    \label{Fx}
\end{align}
where $\theta$ and $\phi$ are the angular sky position and $\psi$ is the polarization angle. 

In GR, for non-precessing and quasi-circular binaries, the inspiral waveform for $+$ and $\times$ polarizations has been calculated in the post-Newtonian (PN) approximation \cite{Blanchet:2013haa}. In the so-called restricted PN approximation (at leading PN order in the amplitudes),  $\tilde{h}_+$ and $\tilde{h}_\times$ can be expressed as
\begin{align}
    \tilde{\bar{h}}_{+} = (1 + \xi^2)A e^{i\Psi}, \label{GR_hp}\\
    \tilde{\bar{h}}_\times = 2\xi A e^{i(\Psi + \pi/2)}\label{GR_hc}.
\end{align}
Here, $\xi = \cos\iota$ where $\iota$ is the inclination angle between the binary's angular momentum vector and the line of sight, $A$ is the amplitude and $\Psi$ is Fourier GW phase in the stationary phase approximation \cite{Bender, Damour:2004bz}.  Higher order PN corrections are also formally  present in $\tilde{\bar{h}}_+$ and $\tilde{\bar{h}}_\times$, however these higher order terms are subdominant. Thus, the main constraining power of beyond-GR parity violation arises from $\tilde{\bar{h}}_{+,\times}$ at Newtonian order, and we can proceed with the restricted PN approximation. Then, the explicit form for the GR detector response function is \cite{Yunes:2010yf}
\be 
\tilde{\bar{h}}(f) = \mathcal{A}f^{-7/6}e^{i(\Psi + \delta\bar{\Psi})},
\ee 
where
\begin{align}
   \mathcal{A} = \sqrt{\frac{5}{96\pi^{4/3}}}\frac{\mathcal{M}^{5/6}}{D_L}\sqrt{F_+^2(1+ \xi^2)^2 + 4 F_\times^2 \xi^2},
\end{align}
and the additional GR contribution to the phase $\delta\Psi$ is given by 
\be 
\delta\Psi =\arctan\frac{F_\times(2\xi)}{F_+(1+\xi^2)}. 
\ee 

Analogously, we would like to characterize the parity-violating GW by writing $\tilde{h}$ in terms of an amplitude and a phase in the following generic way:
\be 
\tilde{h} = 
\tilde{\bar{h}}(1 + \delta\mathcal{A}_A + \delta\mathcal{A}_V)e^{i(\delta\Psi_A + \delta\Psi_V)},\label{PV_strain}
\ee 
where $\delta\mathcal{A}_{A,V}$ and $\delta\Psi_{A,V}$ are the corrections to the amplitude and phase arising from our $\delta\phi_A$ and $\delta\phi_V$ and $\tilde{\bar{h}}$ is the GR value of the response function (obtained replacing Eqs.\ (\ref{GR_hp})-(\ref{GR_hc}) into Eq.\ (\ref{strain})).

To determine the explicit form for the quantities $\delta\mathcal{A}_{A,V}$ and $\delta\Psi_{A,V}$, we first insert the modified expressions for $h_+$ and $h_\times$, Eqs.~\eqref{hp} and~\eqref{hx} into the expression for the response function, Eq.~\eqref{strain}. We then can write the full complex expression as an amplitude and a phase, and expand each to linear order in $\delta\phi_{A,V}$. From this procedure, we obtain:
\begin{align} 
\delta\mathcal{A}_A + \delta\mathcal{A}_V =   f(F_{+,\times}, \xi)\delta\phi_A -  g(F_{+,\times}, \xi)\delta\phi_V \label{dA}
\end{align} 
for the parity-violating amplitude corrections and 
\begin{align}
\delta\Psi_A + \delta\Psi_V =  g(F_{+,\times}, \xi)\delta\phi_A   +  f(F_{+,\times}, \xi)\delta\phi_V
\end{align}
for the parity-violating phase corrections,  where for notational ease, we have defined the functions 
\begin{align}
   f(F_{+,\times}, \xi) &= \frac{ 2(F_+^2 + F_\times^2)\xi(1 + \xi^2)}{4F_\times^2\xi^2 + F_+^2(1 + \xi^2)^2} \label{f}, \\
   g(F_{+,\times}, \xi) &= \frac{F_+F_\times(-1 + \xi^2)^2}{4F_\times^2\xi^2 \label{g}
+ F_+^2(1 + \xi^2)^2}.
\end{align}
The modified response function can then be expressed as 
\begin{align} 
\tilde{h} &= \tilde{\bar{h}}\left[1 + f(F_{+,\times}, \xi)\delta\phi_A - g(F_{+,\times}, \xi)\delta\phi_V \right]\nonumber \\
&\times \exp\left\{i\left[g(F_{+,\times}, \xi)\delta\phi_A + f(F_{+,\times}, \xi)\delta\phi_V\right]\right\}.
\label{tildehfull}
\end{align} 
Note here that when averaging over the entire sky, $g(F_{+,\times}, \xi)$ will vanish due to the cross terms proportional to $F_+F_\times$. However, this is not generically the case if the sky localization of a GW event is known, so we keep these terms for completeness. Furthermore, there are particular angles, e.g., when $F_+ = \xi = 0$ where the linear order parity-violating contribution becomes undefined, in which case one must use the general party-violating expression before Taylor expanding in $\delta\phi_{A,V}$. We show this explicitly in Appendix \ref{App:hFull}. This result for the waveform modification is in agreement with previous results in the literature \cite{Zhao:2019xmm, Yagi:2017zhb}. 

\subsection{Degeneracies with Waveform Parameters}
The parity-violating corrections to the phase, $\delta\phi_V$, could be degenerate with other waveform parameters that represent properties of the system, so let us study this here further. First, if $\delta\phi_V$ is frequency independent, then it is degenerate with the polarization angle, $\psi$, and can be absorbed into the expressions for $F_+$ and $F_\times$, Eqs.~\eqref{Fp} and ~\eqref{Fx}. Note that gravitational waveforms also depend on additional frequency-independent angular parameters such as the coalescence phase, $\phi_c$, but such a parameter affects in the same way the right- and left-handed polarizations of GWs, and hence it will not be degenerate with $\delta\phi_V$. 

We thus notice that we can make a redefinition of $\psi$ such that:
\begin{align}
\psi \rightarrow \hat{\psi} &=\bar{\psi} + \delta\psi \nonumber \\
&= \bar{\psi} + \frac{\delta\phi_V}{2},
\end{align} 
where $\bar{\psi}$ is the true polarization angle and $\delta\psi$ is the correction due to the velocity birefringence. We can then simplify to obtain 
\be 
\delta\mathcal{A}_A = 
 \frac{ 2(F_+(\theta, \phi,\hat{\psi})^2 + F_\times(\theta, \phi, \hat{\psi})^2)\xi(1 + \xi^2)}{(4F_\times(\theta, \phi,\hat{\psi})^2\xi^2 + F_+(\theta, \phi,\hat{\psi})^2(1 + \xi^2)^2} \delta\phi_A
\ee 
and 
\be 
\delta\Psi_A =  \frac{F_+(\theta, \phi, \hat{\psi})F_\times(\theta, \phi, \hat{\psi})(-1 + \xi^2)^2}{4F_\times(\theta, \phi,\hat{\psi})^2\xi^2 + F_+(\theta, \phi,\hat{\psi})^2(1 + \xi^2)^2}\delta\phi_A,
\ee 
with $\delta\mathcal{A}_V = \delta\Psi_V = 0$ because the $\delta\phi_V$ dependence is now in $\delta\psi$. We will denote $\hat{F}_{+,\times} = F_{+,\times}(\theta,\phi, \hat{\psi})$. 

Similarly, if the parameter $\delta\phi_A$ is frequency independent, then it is simultaneously degenerate with the inclination angle $\xi$ and the amplitude of the signal $\mathcal{A}$, which can be seen by the following shifts:
\begin{align}
\xi \rightarrow \hat{\xi} &= \bar{\xi} + \delta\xi \nonumber \\
&= \bar{\xi} + \frac{1}{2}(1-\bar{\xi}^2)\delta\phi_A, \\ 
\mathcal{A} \rightarrow \hat{\mathcal{A}} &= \bar{\mathcal{A}} + \delta\mathcal{A}\nonumber \\
&= \bar{\mathcal{A}} + \bar{\mathcal{A}}\bar\xi\delta\phi_A,
\end{align}
where $\bar{\mathcal{A}}$ and $\bar{\xi}$ are the true values of the amplitude and $\xi$. We emphasize that $\delta\phi_A$ can be absorbed by a simultaneous change in both $\mathcal{A}$ and $\xi$, but it \textit{cannot} be fully absorbed into $\delta\xi$ or $\delta\mathcal{A}$ individually. Notice in particular that the degeneracy with $\mathcal{A}$ is really a degeneracy with the luminosity distance of the source, as the chirp mass affects the amplitude (and phase) of the wave in a  different frequency-dependent way. The full expression for the response function then becomes

\begin{align}
    \tilde{h} &= \hat{F}_+(1 + \hat{\xi}^2)\hat{\mathcal{A}} e^{i\Psi} + i\hat{F_\times} 2\hat{\xi}\hat{\mathcal{A}} e^{i\Psi} 
\\ &=
\hat{\mathcal{A}}\sqrt{\hat{F}_+^2(1+ \hat{\xi})^2 + 4 \hat{F}_\times^2 \hat{\xi}^2}e^{i(\Psi + \delta\hat{\Psi})}, 
\end{align}
where $\delta\hat{\Psi}$ refers to $\delta\Psi$ with $F_{+,\times}$ and $\xi$ replaced with their hatted quantities. Notice that the expression above takes the same form as $\tilde{\bar{h}}$ but is modified due to $\psi$, $\xi$ and $\mathcal{A}$ being modified from their true values. Furthermore, in the scenario in which the observed waveform is only a small portion of the full signal, the amplitude modulations caused by precession in binary systems can mimic the presence of amplitude birefringence\cite{Ng:2023jjt}.

Overall, these degeneracies can lead to complications when measuring birefringence effects. As first noted in \cite{Alexander:2007kv}, if the polarization or inclination angle of a source is uncertain, it can be difficult to separate the effects of velocity or amplitude birefringence from the true values of the source parameters. Electromagnetic counterparts, such as those from GW170817, can help constrain polarization and inclination angles and eliminate uncertainty surrounding birefringence effects. Nevertheless, if $\delta \phi_{A,V}$ do depend on frequency, and the GW detection covers a wide range of frequencies (e.g.,\ the signal is generated by a light compact binary event), then these degeneracies are expected to be broken. Notice that even for a single detector in cases where the GR amplitude and phase may coincidentally contain PN terms with the same frequency evolution as $\delta \phi_{A,V}$, there most likely will not be a degeneracy since the GR terms will depend on the same binary parameters that will affect other, different, frequency-dependent terms and cannot be chosen freely to reabsorb the effects of $\delta \phi_{A,V}$. For multiple detectors the degeneracy will be even less likely since $\delta \phi_{A,V}$ additionally will affect right- and left-handed polarizations differently that can be independently measured with detectors oriented differently.

\subsection{Mapping to PPE}
Now that we have obtained the parity-violating modification in a standard form in terms of the waveform  response, we can map it to the ppE framework \cite{PPE, Sampson:2013wia, cornishsampson, Chatziioannou:2012rf}. The simplest ppE waveform in the frequency domain can be written as 
\be 
\tilde{h}_\ppe = \tilde{\bar{h}}\left(1 + \alpha_\ppe u^{a_\ppe} \right)e^{i \beta_\ppe u^{b_\text{ppE}}},
\ee 
where $\alpha_\ppe, \beta_\ppe, a_\ppe$ and $b_\ppe$ 
are dimensionless ppE parameters, which can be mapped to various theories, and $u = \pi \mathcal{M}f$, where $f$ is the detected GW frequency and $\mathcal{M}$ is the GW chirp mass. The latter encodes information about the binary component masses, and is the main effective parameter determining the GW morphology during the inspiral. 

Let us now calculate the four parameters $\alpha_\ppe,\, \beta_\ppe, \,a_\ppe$, and $b_\ppe$ for the parity-violating model discussed in the previous sections. We can  write the expression for $\tilde{h}$ by expanding out $\delta\phi_A$ and $\delta\phi_V$ explicitly in Eq.~\eqref{tildehfull}. Then we can  map the amplitude and phase corrections in a ppE form as
\be
\tilde{h} = \tilde{\bar{h}} \left(1 + \sum_{a_\ppe} u^{a_\ppe }\alpha_{a_\ppe}^{\ppe}\right)e^{i\sum_{b_\ppe}u^{b_\ppe}\beta_{b_\ppe}^{\text{ppE}}},
\ee 
where we can read off the ppE parameters to be 
\be
a_\ppe = b_\ppe = n, m
\ee
and in the case of $a_\ppe = b_\ppe=n$, the associated coefficients are:
\begin{align}
\alpha^\ppe_{n}  &= \left[\frac{2(1+z)}{\mathcal{M}\Lambda_\pv }\right]^{n}\frac{f(F_{+,\times},\xi)}{2} \left[z_n\alpha_{n_0} + \Lambda_\pv D_{n+1}\beta_{n_0}\right],\\
\beta^\ppe_{n} &= \left[\frac{2(1+z)}{\mathcal{M}\Lambda_\pv }\right]^{n}\frac{g(F_{+,\times},\xi)}{2} \left[z_n\alpha_{n_0}, \Lambda_\pv D_{n+1}\beta_{n_0}\right],
\end{align}
while in the case of $a_\ppe = b_\ppe=m$ we have
\begin{align}
\alpha^\ppe_{m} &=- \left[\frac{2(1+z)}{\mathcal{M}\Lambda_\pv }\right]^{m}\frac{g(F_{+,\times},\xi)}{2} \left[z_m\gamma_{m_0} + \Lambda_\pv D_{m+1}\delta_{m_0}\right],\\
\beta^\ppe_{m} &= \left[\frac{2(1+z)}{\mathcal{M}\Lambda_\pv }\right]^{m}\frac{f(F_{+,\times},\xi)}{2} \left[z_n\gamma_{n_0} + \Lambda_\pv D_{n+1}\delta_{n_0}\right].
\end{align}

These general expressions appear quite cumbersome, however they greatly simplify when considering specific parity violating theories. We can see explicitly as well that $\alpha^\ppe_{a_\ppe}$ and $\beta^\ppe_{b_\ppe}$ are dimensionless quantities, as all dimensional quantities appear with the correct power of $\Lambda_\pv$ to render the full expression dimensionless.

Table \ref{tab:PPEMap} shows the specific ppE parameters and corresponding post-Newtonian order for each theory. Recall that an $N$-PN order corresponds to a contribution of $\mathcal{O}(u/c)^{2N/3}$ compared to the leading order GR term in the waveform. For that given order, this ppE formulation gives corrections  $a_\text{ppE} = \frac{2N}{3}$ and $b_\text{ppE} = \frac{2N-5}{3}$.
\begin{widetext}
\begin{center}
\begin{table}[h]
    \centering
    \begin{tabular}{|c|c|c|c|c|c|c|} \hline 
        Theory &$a_\text{ppE}$ & $\alpha^\text{ppE}$ & PN & $b_\text{ppE}$ & $\beta^\text{ppE}$ & PN \\ \hline \hline 
       Chern-Simons \cite{Alexander:2007kv, Yunes:2010yf, Alexander:2017jmt} & 1 & $f(F_{+,\times}, \xi)\frac{z_1(1+z)}{\mathcal{M}}\frac{\alpha_{1_0}}{\Lambda_\pv}$  & 1.5  & 1  & $g(F_{+,\times}, \xi)\frac{z_1(1+z)}{\mathcal{M}}\frac{\alpha_{1_0}}{\Lambda_\pv}$ &  4 \\\hline 
        Ghost-Free Scalar Tensor \cite{Nishizawa:2018srh, Zhao:2019xmm,  Qiao:2021fwi} &1
 & $f(F_{+,\times},\xi)\frac{(1+z)}{\mathcal{M}}\left(z_1\frac{\alpha_{1_0}}{\Lambda_\pv} + \beta_{1_0}D_2\right)$& 1.5 & 1 & $g(F_{+,\times},\xi)\frac{(1+z)}{\mathcal{M}}\left(z_1\frac{\alpha_{1_0}}{\Lambda_\pv} + \beta_{1_0}D_2\right)$  & 4 \\
 {} & 2 &$-2 g(F_{+,\times},\xi)\frac{D_3(1+z)}{\mathcal{M}^2}\frac{\delta_{2_0}}{\Lambda_\pv}$& 3&2 &$2 f(F_{+,\times},\xi)\frac{D_3(1+z)}{\mathcal{M}^2}\frac{\delta_{2_0}}{\Lambda_\pv}$&5.5 \\ \hline 
        Symmetric Teleparallel I \cite{Conroy:2019ibo}&  0 &  $-\frac{g(F_{+,\times},\xi)}{2}z_0\gamma_{n_0}$ &  0& 0 & $\frac{f(F_{+,\times},\xi)}{2} z_0\gamma_{n_0}$ & 2.5 \\ \hline 
        Symmetric Teleparallel II \cite{Conroy:2019ibo}& 1 & $f(F_{+,\times},\xi)\frac{(1+z)}{\mathcal{M}}\left(z\frac{\alpha_{1_0}}{\Lambda_\pv} + \beta_{1_0}D_2\right)$ & 1.5 & 2 &$g(F_{+,\times},\xi)\frac{(1+z)}{\mathcal{M}}\left(z\frac{\alpha_{1_0}}{\Lambda_\pv} + \beta_{1_0}D_2\right)$  & 4\\ 
        {} & 2 & $-2 g(F_{+,\times},\xi)\frac{D_3(1+z)}{\mathcal{M}^2}\frac{\delta_{2_0}}{\Lambda_\pv}$& 3 &2 & $2 f(F_{+,\times},\xi)\frac{D_3(1+z)}{\mathcal{M}^2}\frac{\delta_{2_0}}{\Lambda_\pv}$& 5.5\\ \hline 
        Horava-Lifshitz \cite{Zhu:2013fja}& 2 & $-2 g(F_{+,\times},\xi)\frac{D_3(1+z)}{\mathcal{M}^2}\frac{\delta_{2_0}}{\Lambda_\pv}$ & 3 & 2 &$2 f(F_{+,\times},\xi)\frac{D_3(1+z)}{\mathcal{M}^2}\frac{\delta_{2_0}}{\Lambda_\pv}$ &    5.5
    \\ 
    {} & 4 & $-8g(F_{+,\times},\xi)\frac{D_5}{\mathcal{M}^4}\frac{\delta_{4_0}}{\Lambda_\pv^3}$  & 6 & 4 & $8f(F_{+,\times},\xi)\frac{D_5}{\mathcal{M}^4}\frac{\delta_{4_0}}{\Lambda_\pv^3}$ & 8.5\\ \hline
    \end{tabular}
    \caption{Parametrized post-Einsteinian mapping of a given detector response for parity-violating theories. }
    \label{tab:PPEMap}
\end{table}
\end{center}
\end{widetext}

In addition to the mapping shown above in terms of the detector response $\tilde{h}$, it is also useful to consider a ppE mapping at the level of $h_+$ and $h_\times$. This is particularly useful if one wishes to stack constraints from many events without needing to be concerned with the detector response and sky localization parameters for each, as well as if there is one event with multiple detectors. In this case, we parametrize separately the modifications in $h_+$ and $h_\times$ as 
\begin{align}
    \tilde{h}_+ &= \tilde{\bar{h}}_+(1 + \delta \mathcal{A}_+)e^{i\delta\Psi_+},\\
    \tilde{h}_\times &= \tilde{\bar{h}}_\times(1 + \delta \mathcal{A}_\times)e^{i\delta\Psi_\times}.
\end{align}
Plugging in the expressions for $\bar{h}_{+,\times}$, Eqs.~\eqref{GR_hp} and ~\eqref{GR_hc}, we find 
\begin{align}
\tilde{h}_{+,\times} &= \tilde{\bar{h}}_{+,\times}\Big[1 + \xi_{+,\times}(\xi)\delta\phi_A \Big]\exp\Big[i \xi_{+,\times}(\xi)\delta\phi_V\Big],
\end{align}
where we define 
\begin{align}
\xi_+(\xi) &= \frac{2\xi}{(1+\xi^2)},\\
\xi_\times(\xi) &= \frac{(1 + \xi)^2}{2\xi}.
\end{align}

Then, in the same way as above, we can write this expression in terms of ppE parameters as:
\begin{align}
a_\ppe &= n,\\
b_\ppe &= m,\\
\alpha^\ppe_{n}  &= \frac{\xi_{+,\times}(\xi)}{2}\left( \frac{2}{\mathcal{M}\Lambda_\pv }\right)^{n}\nonumber \\
   & \times \left[z_n\alpha_{n_0} + \Lambda_\pv D_{n+1}\beta_{n_0}\right] ,\\
\beta^\ppe_{m} &=\frac{\xi_{+,\times}(\xi)}{2}\left( \frac{2}{\mathcal{M}\Lambda_\pv }\right)^{m}\nonumber \\
   & \times \left[z_m\gamma_{m_0}+ \Lambda_\pv D_{m+1}\delta_{m_0}\right]. 
\end{align}
Again, we see that $\alpha^\ppe_{a_\ppe}$ and $\beta^\ppe_{b_\ppe}$ are dimensionless quantities as required.
Notice that the ppE parameters for $h_+$ and $h_\times$ differ by a $\xi$ dependent prefactor. One may be concerned about the factor of $\xi^{-1}$ in $\xi_\times(\xi)$ when $\xi=0$. However, we note the apparent divergence cancels out with the GR factor in $\bar{h}_\times$ and as along as deviations from GR remain small there is no issue. 

\section{Constraints}
\label{Constraints}
Full data analysis with the parametrization introduced in this work will be necessary to rigorously constrain the new parity-violating parameters. However, as a first step, we can consider initial constraints based on previously existing work in the literature. In this section, we consider both the velocity constraints from the GW170817/GRB170817 coincident event and birefringence specific constraints in the literature from binary black hole events. We note that the LVK analysis on modified dispersion relations and propagation effects does not include birefringence, and, thus, we will not include it for the purposes of this paper. 

\subsection{Propagation Speed}
The coincident gravitational wave/gamma ray burst event from binary neutron star merger, GW170817 has provided a tight constraint on the speed of GWs, $c_T$, compared to the speed of light, $c$. We have \cite{LIGOScientific:2017zic}
\be 
-7 \times 10^{-16} < 1 - c_T <3\times 10^{-15}.
\label{eq:gwspeed}
\ee 
This observation immediately ruled out many beyond-GR theories that induce a modification to the GW speed \cite{Baker:2017hug,Creminelli:2017sry, Sakstein:2017xjx, Ezquiaga:2017ekz, Wang:2017rpx}. We can map this constraint to our parametrization using our expression for the modified group velocity, Eq.\ \eqref{Eq:vg}, assuming that both polarization still propagate coherently in time. Notice that when $m=0$, the parity-violating contribution to $v_g$ vanishes, although this model will still induce phase velocity modifications according to Eq.\ (\ref{Eq:vp}) and it will hence affect the overall phase of the signal (see related discussions in \cite{Ezquiaga:2022nak}). 
If we then consider $m\not=0$, we find that the $\gamma_m$ term in $v_g$  will be suppressed by a factor of $\mathcal{H}/\Lambda_{PV}$ compared to the $\delta_m$ term, and thus we will focus on the latter, bigger, term. We thus approximate:
\be 
v_g -1\approx \lambda_\rl m\left(\frac{k}{a}\right)^{m-1}\frac{\delta_m}{\Lambda_\pv^{m-1}}.
\ee 
Taking the weaker constraint from Eq.~\eqref{eq:gwspeed}, we find that 
\be 
\left|\left(\frac{k}{a}\right)^{m-1}\frac{\delta_m}{\Lambda_\pv^{m-1}}\right| < 3 \times 10^{-15}.
\ee 
The case of $m=2$ is especially important since it indeed corresponds to the phase modification that appears in parity violating ghost-free scalar-tensor theory, symmetric teleparallel equivalent of GR and Horava-Lifshitz theory, as can be seen in \ref{tab:TheoryMap}. Considering a frequency at merger of $k/a \sim 10^3$ Hz, we find 
\be 
\left|\frac{\delta_2}{\Lambda_\pv} \right| \lesssim 10^{-9} \text{ m}. \label{cT_const}
\ee 
If we take $\delta_2$ to be an $\mathcal{O}(1)$ quantity, we can invert this constraint into a lower bound on the cutoff scale, $\Lambda_\pv \gtrsim  10^2$ eV. Note that in obtaining this bound we convert from Eq.\ (\ref{cT_const}) to natural units ($c=\hbar=1$) to make contact with the usual conventions for effective field theories in particle physics. 

\subsection{Polarization and Phase}
We will now consider how previous analyses regarding the polarization of binary populations and phase distortions translate into our parametrization. 

In \cite{Okounkova:2021xjv} the authors performed an initial study to constrain amplitude birefringence with binary black hole mergers from the GWTC-2 data set. The authors considered a toy model for amplitude birefringence, taking $h_\rl  = \bar{h}_\rl e^{\lambda_\rl  \kappa D_C}$, where  $\kappa$ is known as the opacity parameter and $D_C$ is the comoving distance. Because one polarization is exponentially enhanced with respect to the other one, if birefringence is present it may induce an overall preference for mostly right or left-handed polarizations in the total GW population. In turn, this would translate into a preference for mostly face-on or face-off binary systems, which is not observed in GWTC-2 and hence sets a bound in $\kappa$.
In this study, the authors assumed all merger events were at a common average comoving distance of $D_C \simeq 1.23$ Gpc (equivalently, $z=0.3$) and that the birefringence effect was frequency independent for simplicity. The constraint on $\kappa$ was found to be 
\be 
\left| \kappa \right|\lesssim 0.74 \text{ Gpc}^{-1}.
\ee

Recall that per the discussion in Sec.~\ref{sec:genparam}, from a theoretical standpoint, parity-violating effects do not enter the propagation equation with even powers of $k$ and thus a $k$-independent amplitude effect does not actually correspond to any parity-violating theory. However, in order to make a direct comparison to \cite{Okounkova:2021xjv}, we will consider a constraint based on $n=0$ in Eq.\ \eqref{h_Eq}. This allows one to translate the opacity constraint directly to a joint constraint on $\alpha_{0_0}$ and $\beta_{0_0}$. Taking the small redshift limit such that $D_C \approx D_1 \approx z/H_0$ and recalling that $z_0 = \ln(1+z)$, we find that 
\be 
\left| \alpha_{0_0} H_0 + \Lambda_\pv \beta_{0_0} \right|\lesssim 1.5 \text{ Gpc}^{-1}.
\ee 

The relevant constraint in a theoretical context is in fact the $n=1$ term, which corresponds to Chern-Simons gravity, ghost-free scalar tensor gravity and symmetric teleparallel gravity. To extend this constraint to include frequency dependence, we consider a representative frequency value of $f\sim 100$ Hz, where current detectors are the most sensitive. From this, we obtain the following order-of-magnitude estimation
\be 
\left|\frac{\alpha_{1_0}}{\Lambda_\pv}H_0 + \beta_{1_0} \right|\lesssim \mathcal{O}(10^{-19}), \label{AmpBire_Const}
\ee 
where all quantities are now dimensionless. We again notice that the $\alpha$ term is suppressed by a factor of $H_0/\Lambda_\pv$ compared to the $\beta$ term. If both parameters are expected to be of the same order, then Eq.\ (\ref{AmpBire_Const}) translates into a constraint in $\beta$ only, $|\beta_{1_0}|\lesssim \mathcal{O}(10^{-19})$. 

Repeating the data analysis of \cite{Okounkova:2021xjv} while actually taking into account the frequency dependence in amplitude birefringence will lead to an overall improvement on the constraint since, in that case, amplitude birefringence will not be degenerate with the binary inclination. In \cite{Wang:2020cub} the authors use the GWTC-1 binary catalog and constrain amplitude birefringence including a linear frequency dependence. The authors obtain a constraint on the scale of parity violation at $100$ Hz, which is equivalent to our parameter in the low-redshift approximation $|\beta_{1_0}|\lesssim 5\times 10^{-19}$.

Furthermore, it was recently shown that using the complete GWTC-3 binary catalog leads to an improvement of several orders of magnitude in the amplitude birefringence constraint \cite{Ng:2023jjt}. The authors parametrize the effect as $\delta \phi_A=\kappa D_C (f/100 \textrm{Hz})$. This specific frequency and time dependence is exactly encoded into the parameter $\beta_{1_0}$ (c.f.\ Eq.\ (\ref{Amp_Bire_Linear}) with $D_2(1+z)=D_C$), then without making any small-redshift approximation we directly translate the constraints from \cite{Ng:2023jjt} into:
\begin{equation}
    |\beta_{1_0}|< 0.7\times 10^{-20}
\end{equation}
at $90\%$ confidence level.

In addition to the constraints on amplitude birefringence, work has been done to study velocity birefringence with LVK data. The authors in \cite{Zhao:2022pun} performed an analysis of the LVK O3 events from the GWTC-2 and GWTC-3 catalogs. The authors considered a modified dispersion relation of the form $\omega^2 = k^2 \pm 2\zeta k^3$, and found a bound on the parameter $\zeta$ such that $|\zeta| \lesssim \mathcal{O}(10^{-17} \text{ m})$. Notice the authors consider $\zeta$ to be a constant factor, differing from our setup in which we assume the parity violating parameters generically can vary with time. Nevertheless, we can map this constraint to our $\delta_{2_0}$ such that we obtain 
\be 
\left| \frac{\delta_{2_0}}{\Lambda_\pv} \right|\lesssim \mathcal{O}(10^{-16}) \text{ m}.
\ee 
If $\delta_{2_0} \sim \mathcal{O}(1)$ then $\Lambda_\pv \lesssim 0.5 \text{ GeV}$. Note that, again, we do not consider $\gamma_{2_0}$ because it is suppressed by an additional factor of $\mathcal{H}/\Lambda_\pv$ compared to $\delta_{2_0}$. Here we see that velocity birefringence leads to several orders of magnitude tighter constraints than those from the overall GW speed in Eq.\ (\ref{cT_const}). This shows that future  analyses that incorporate frequency-dependent distortions of the waveform will be the most promising way of improving  parity-violation constraints for models in which both effects are present. Nevertheless, some theories like Chern-Simons gravity are only expected to exhibit amplitude birefringence. 

Further analysis will need to be done in order to fully explore the parameter space of parity-violating theories with our newly introduced parametrization; however, these existing constraints provide some initial insight into the types of bounds we might expect to obtain with new data analysis of LVK merger events. 

\section{Discussion and Conclusions}
\label{Sec:discussion}

In this work we have introduced a new parametrization scheme to describe GW propagation in parity-violating extensions to GR, which we argue is the most generic modification of the GW propagation equation to describe parity violation. We have shown how these generic modifications to the propagation equations impact the right and left-handed circular polarization modes, leading to both amplitude and phase modifications in the waveform. Known parity-violating theories in the literature can be described using this parametrization, which we show explicitly via the mapping in Table \ref{tab:TheoryMap}. Furthermore, the modified polarization modes in the circular polarization basis translate easily to a ppE waveform template, both at the level of the detector response function $h$ and at the level of the individual polarization states  $h_+$ and $h_\times$. Lastly, we have shown how current constraints in the literature map onto our parametrization. We have discussed constraints on the modified dispersion relation from the GW170817/GRB170817 coincident event, as well as current bounds from studies of amplitude and velocity birefringence from binary black hole mergers in the GWTC-2 and GWTC-3 catalogs.  

There are a variety of pathways forward towards future work. On the theory side, in the above work, we have limited ourselves to the assumption that if the birefringence parameters $\alpha, \beta, \gamma,$ and $\delta$ are small, and that if they vary with time, they can be well approximated by their present day values, i.e., that they vary slowly compared to the expansion of the universe. One could drop this assumption and consider the parametrized modifications arising from a higher-order expansion in the birefringence parameters or even considering a universal time evolution profile. We also note that we have not included parity-violating generation effects in this current study. In general, parity violation impacts the generation of GWs by leading to a modification in the chirping rate \cite{Yagi:2011xp, Carson:2020iik, Shiralilou:2020gah, Shiralilou:2021mfl, Okounkova:2019dfo, Okounkova:2022grv}. Generally, this enters at a high PN order, but can be enhanced by considering eccentric, highly spinning binaries \cite{Alexander:2017jmt,Loutrel:2022tbk}. A future realistic test of parity violation would take into account both generation and propagation effects in parity-violating theories for the above-mentioned observational probes. 

 Most importantly, further work should be done to investigate how this parametrization scheme can be used to place meaningful bounds on parity violation from binary black hole events, binary neutron star events, and the stochastic GW background. This includes data analysis from the LVK binary black hole and binary neutron star events detected so far as well as forecasting constraints from future observations. The observations by the LVK O4/O5 observing runs, LISA \cite{2017arXiv170200786A} and potential third-generation experiments such as the proposed Einstein Telescope \cite{Maggiore:2019uih} and Cosmic Explorer \cite{Evans:2021gyd} will no doubt be increasingly sensitive probes to the scale of amplitude and velocity birefringence in GW propagation.

\section*{Acknowledgements}
We thank Stephon Alexander, Will Farr, Wayne Hu, and Maximiliano Isi for useful comments and discussions. The work of L.J.\ is supported by the Kavli Institute for Cosmological Physics at the University of Chicago through an endowment from the Kavli Foundation and its founder Fred Kavli. M.L.\ was supported by the Innovative Theory Cosmology Fellowship at Columbia University. N.~Y.~acknowledges support from NSF Grant No. PHY-2207650 and from the Simons Foundation through Award No. 896696.
\appendix

\section{Summary of Parity-Violating Theories}
\label{App:Theories}
In this appendix we discuss specifics of the parity-violating theories to which our parametrization maps. 

\subsection{Chern-Simons Gravity}
The most well-studied parity-violating modified gravity theory is Chern-Simons gravity. It is characterized by the addition of a gravitational Chern-Simons term to the Einstein-Hilbert action such that 
\be 
S = \int d^4 x \sqrt{-g} \left(R + \frac{\alpha^\cs}{4\kappa}\vartheta *RR\right), 
\ee 
where $\kappa = (16\pi)^{-1}$, $\alpha^\cs$ is a coupling parameter, the pseudo-scalar field, $\vartheta$ is coupled to the Pontryagin density of the spacetime, defined by 
\be 
*R^a{}_b{}^{cd}R^b{}_{acd}, 
\ee 
where the Hodge dual to the Riemann tensor is 
\be 
*R^a{}_b{}^{cd} = \frac{1}{2}\epsilon^{cdef}R^a{}_{bef}, 
\ee 
with $\epsilon^{cdef}$ the totally antisymmetric Levi-Civita tensor. The propagation of gravitational waves in Chern-Simons gravity on a cosmological background has been well studied in \cite{Alexander:2007kv, Yunes:2010yf, Alexander:2017jmt, Yagi:2017zhb} so we will present only a schematic outline of the computation here. The linearized equations of motion are 
\be 
    \Box h^j{}_i = -\frac{1}{a^2}\epsilon^{pjk}\left[(\vartheta^{''} - 2\mathcal{H}\vartheta')\partial_p h_{ki}' + \vartheta' \partial_p\Box h_{ki}\right],
\ee 
which after taking a plane-wave ansatz, leads to a modified dispersion relation:
\begin{align}
&i\phi(\eta)^{''}  + \phi(\eta)'^2 - k^2 \nonumber \\
&= -2i\phi(\eta)' \left(\mathcal{H} -  \frac{4\alpha^\cs \lambda_\rl }{\kappa 2a^2} k\vartheta^{''}\right)\left(1 + \frac{4\alpha^\cs \lambda_\rl }{\kappa a^2}k\vartheta' \right).
\end{align} 
Linearizing the dispersion relation and using the equation of motion for the pseudo-scalar field to obtain $\ddot{\vartheta} = -2H\dot{\vartheta}$, we obtain 
\be 
\delta\phi = -i\lambda_\rl \frac{k}{2}\frac{4\alpha^\cs \dot{\vartheta}_0}{\kappa}z.
\ee 
Lastly, notice that the quantity $\alpha \dot{\vartheta}/\kappa$ has units of length (in geometric units). To make this dimensionless, we redefine coupling, $\tilde{\alpha}^\cs = \alpha^\cs \Lambda_\pv$, and make the replacement $\alpha^\cs \rightarrow \tilde{\alpha}^\cs$ such that the entire expression is dimensionless, and we obtain the result in Table \ref{tab:TheoryMap}.

Generalized Chern-Simons gravity \cite{Nojiri:2019nar}(also denoted as Chern-Simons-Axion-Einstein gravity)  generalizes the scalar field coupling to the Chern-Simons to a generic function of the scalar, $U(\phi)$. From this, one can obtain a modified dispersion relation such that at this order of study we have
\be 
\omega^2 = k^2 - i\lambda_\rl  k U_0 H, 
\ee 
and we find the same result for $\delta\phi$ as in the usual Chern-Simons case, with a slightly different notation. 
\subsection{Ghost-Free Scalar Tensor Gravity}
Parity-violating ghost-free scalar tensor theories allow extensions beyond Chern-Simons gravity by considering higher derivative terms of the scalar field. In \cite{Nishizawa:2018srh} the following contributions to the Lagrangian are considered: 
\begin{align}
L_1 & = \epsilon^{\mu\nu\alpha\beta}R_{\alpha\beta\rho\sigma}R_{\mu\nu}{}^\rho{}_\lambda\phi^\sigma\phi^\lambda,\\
L_2 &= \epsilon^{\mu\nu\alpha\beta}R_{\alpha\beta\rho\sigma}R_{\mu\lambda}{}^{\rho\sigma}\phi_\nu\phi^\lambda, \\
L_3 &= \epsilon^{\mu\nu\alpha\beta}R_{\alpha\beta\rho\sigma} R^\sigma{}_\nu \phi^\rho \phi_\mu,\\
L_4 &= \phi_\lambda\phi^\lambda\epsilon^{\mu\nu\rho\sigma}R^{\alpha\beta}{}_{\mu\nu}, 
\end{align}
where $\phi$ is the scalar field. The equations of motion can be found to be 
\begin{align} 
h_{ij}^{''} &+ 2\mathcal{H} h_{ij}' - \partial^2h_{ij} \nonumber \\
&+ \frac{1}{a\Lambda}\epsilon_{ilk}\partial_l\left[ \alpha^\gf h_{jk}^{''} + (\mathcal{H}\alpha^\gf+ (\alpha^\gf)')h_{jk}' - \beta^\gf \partial^2h_{jk}\right] = 0,
\end{align} 
with $\Lambda$ the cutoff scale of the theory and $\alpha^\gf, \beta^\gf$ functions of $\dot{\phi}$, leading to 
\begin{align}
    \delta\phi &=   \frac{ik\lambda_\rl }{2\Lambda}\left[\dot{\alpha}^\gf_0(1 +z) D_2 - \alpha^\gf_0 z\right] \nonumber \\
    &+  \frac{\lambda_\rl k^2}{2\Lambda}(\beta^\gf_0 - \alpha^\gf_0)(1+z)^2D_3.
\end{align}
As in the CS case above, we will rescale $\dot{\alpha}^\gf \rightarrow \dot{\tilde{\alpha}}^\gf= \dot{\alpha}^\gf/\Lambda_\pv$ to make the parity violating parameter dimensionless.

\subsection{Symmetric Teleparallel Equivalent of General Relativity}

 In this section we follow the discussion in \cite{Conroy:2019ibo}. The symmetric teleparallel equivalent of GR is a non-metric theory defined with respect to the non-metricity tensor: 
\begin{align}
Q_{abc} &= \nabla_ag_{bc},\\
Q_a &= g^{bc}Q_{abc},\\
\tilde{Q}_c &= g^{ab}Q_{abc}, 
\end{align}
with an action given by 
\begin{align}
S_{QGR} &= -\frac{1}{2\kappa}\int d^4x \sqrt{-g}\Bigg[-\frac{1}{4}Q_{abc}Q^{abc} \nonumber \\&+ \frac{1}{2}Q_{abc}Q^{bac} + \frac{1}{4}Q_aQ^a - \frac{1}{2}Q_a\tilde{Q}^a\Bigg]. 
\end{align} 
One can construct parity-violating extensions to this at both second and higher order. Considering at most second order in derivatives, the parity-violating contributions are
\begin{align}
L_{a3} &= \epsilon^{abcd}\phi_c\phi^fQ_{abe}Q_{fd}{}^e,\\
L_{a7} &= \epsilon^{abcd}\phi_f\phi^f Q_{abe}Q_{cd}{}^e, 
\end{align}
where $\alpha$ is a function of $\dot{\phi}$, an auxiliary scalar field. The equations of motion become:
\be   
\left(h_{ij}^{''} + 2\mathcal{H} h_{ij}' - \partial^2 h_{ij}\right) - 4\mathcal{H}\alpha\epsilon_{kl(i}\partial^kh^l_{j)} = 0,
\ee
which can be converted into a dispersion relation
\be 
i\phi^{''} + 2i\mathcal{H}\phi' + \phi'^2 - k^2 + 4 k \alpha \mathcal{H}\lambda_\rl = 0.
\ee 
to find 
\be 
\delta\phi = -2\lambda_\rl \alpha_0 \ln(1+z).
\ee 
By including higher derivative terms, one can obtain additional contributions to the action:
\be 
S_{PV} = \frac{1}{2\kappa} \int d^4 x a^3 \left(\frac{\beta_1(t)}{a^3\Lambda}\mathcal{L}_{PV1} + \frac{\beta_2(t)}{a\Lambda}\mathcal{L}_{PV2} + \frac{\beta_3(t)}{a\Lambda}\mathcal{L}_{PV3}\right),
\ee 
where 
\begin{align}
    \mathcal{L}_{PV1} &= \epsilon^{ijk}\partial^2 h_j{}^l \partial_i h_{kl},\\
    \mathcal{L}_{PV2} &= 2H\epsilon^{ijk} \dot{h}_j{}^l\partial_i h_{kl},\\
    \mathcal{L}_{PV3} &= \epsilon^{ijk}\dot{h}_j{}^l \partial_i\dot{h}_{kl}
\end{align}
and 
\begin{align}
    \tilde{\beta}_1 &= (\beta_2' \mathcal{H} + \beta_2\mathcal{H}') + 3(\beta_3'\mathcal{H} + \beta_3\mathcal{H}') + \beta_3\mathcal{H}^2,\\
    \tilde{\beta}_2 &= \beta_2' + 3\mathcal{H}\beta_2. 
\end{align}
Keeping only the relevant correction terms, we obtain 
\begin{align}
\phi^{''} &+ \phi'^2 + i\phi' \left[2\mathcal{H} + \lambda_\rl \frac{k}{a\Lambda}\beta_2' + \lambda_\rl \frac{k}{a\Lambda}\mathcal{H}(3\beta_2 - 2\beta_3)\right]\nonumber \\
&- k^2\left[1 + \lambda_\rl \frac{k}{a\Lambda}\left(\beta_1- \beta_3\right)\right] =0,
\end{align}
and 
\begin{align} 
\delta\phi &=  -\frac{i\lambda_\rl k}{2\Lambda}\left[\dot{\beta}_{2_0}(1 + z)D_2 + (3{\beta_2}_0 - 3{\beta_3}_0 )z\right]\nonumber \\
&+  \frac{\lambda_\rl }{2\Lambda}k^2(1+z)^2 ({\beta_1}_0 - {\beta_2}_0)D_3. 
\end{align}
Lastly, we again rescale $\dot{\beta}_2 \rightarrow \dot{\tilde{\beta}}_2 = \dot{\beta}_2 /\Lambda_\pv$.

\subsection{Horava-Lifshitz Gravity}
The parity-violating extension of Horava-Lifshitz gravity can be found in \cite{Zhu:2013fja} and is characterized by adding parity-violating terms 
\be 
\mathcal{L}_{PV} =\frac{\alpha_2\epsilon^{ijk}R_{il}\Delta_j R^l_k }{M_*^3}+ \frac{\alpha_1\omega_3(\Gamma)}{M_*}
\ee 
to the usual Horava-Lifshitz action, where $\alpha_1$ and $\alpha_2$ are constants, $M_*$ is the cutoff scale of the theory and $\omega_3$ is the usual gravitational Chern-Simons term. The field equations become: 
\begin{align}
h_{ij}^{''} &+ 2\mathcal{H} h_{ij}' - \partial^2 h_{ij} + h_{ij} \nonumber \\
&+ \epsilon_i{}^{lk}\left(\frac{2\alpha_1}{M_*a} + \frac{\alpha_2}{M_*^3 a^3}\partial^2\right)\partial_l\partial^2h_{jk} = 0.
\end{align}
We can then obtain the dispersion relation
\be 
i\phi^{''} + \phi'^2 + 2 i \mathcal{H}\phi' - k^2  + k^3\lambda_\rl \left(\frac{2\alpha_1}{M_*a} - \frac{\alpha_2}{M_*^3 a^3}k^2\right)=0,
\ee 
and thus 
\be 
\delta\phi = -\frac{\alpha_1\lambda_\rl }{2M_*} k^2 (1+z)^2 D_3 + \frac{\lambda_\rl \alpha_2}{2M_*^3} k^4(1 + z)^4 D_5.
\ee 

\section{General Form of Parity-Violating Corrections to Detector Response}
\label{App:hFull}
In this appendix we derive the generalized form of the detector response given in Eqs.\ \eqref{dA}-\eqref{g}. 
Using the modifications of $h_+$ and $h_\times$ given by Eqs.\ \eqref{hp}-\eqref{hx}, we substitute into the detector response function to obtain
\begin{align}
\tilde{h} = A\delta Ae^{i(\Psi + \delta\Psi)}
\end{align}
where $A$ and $\Psi$ correspond to the GR amplitude and Fourier GW phase introduced in Eq.\ (\ref{GR_hp})-(\ref{GR_hc}) in the stationary phase approximation respectively. The parity-violating modifications are generally included in the terms $\delta A$ and $\delta \Psi$, which are given by:
\begin{widetext}
\be 
\delta A = \sqrt{(F_+(1 + \xi^2)+F_+\delta\phi_A 2\xi-F_\times\delta\phi_V (1 + \xi^2))^2} \\ \overline{+ (F_+\delta\phi_V 2\xi + F_\times2\xi + F_\times\delta\phi_A (1 + \xi^2))^2}, 
\ee
\end{widetext}
and  
\begin{align}
\delta\Psi &= \arctan{\frac{(F_+\delta\phi_V 2\xi + F_\times2\xi + F_\times\delta\phi_A (1 + \xi^2))}{(F_+(1 + \xi^2)+F_+\delta\phi_A 2\xi-F_\times\delta\phi_V (1 + \xi^2))}}
\end{align}
Taylor expanding (B2) and (B3) to first order assuming small $\delta\phi_A$ and $\delta\phi_V$ results in Eqs.\ \eqref{dA}-\eqref{g}. 

In the specific case where $F_+ = \xi=0$ (and hence $F_\times=1$), the functions $f(F_{+,\times}, \xi)$ and $g(F_{+,\times}, \xi)$ from Eqs.\ \eqref{f}-\eqref{g} are undefined, so we instead have to use Eqs.\ (B2) and (B3) and obtain
\begin{align}
\delta A &= \sqrt{(\delta\phi_V)^2 + (\delta\phi_A)^2} \\
\delta\Psi &= \arctan\left(\frac{-\delta\phi_A}{\delta\phi_V}\right).
\end{align}
\bibliography{master}

 \end{document}